\documentclass[a4paper,twoside,10pt]{article}
\usepackage{graphicx,publaob}

\def\papertype{Contributed paper}
\begin{document}

\title{SPECTROSCOPIC OBSERVATIONS OF HIGH PROPER MOTION DA WHITE DWARFS}

\authors{E. ARAZIMOV\'A$^1$, A. KAWKA$^1$ \& S. VENNES$^2$}

\address{$^1$Astronomick\'y \'ustav AV \v{C}R, CZ-251 65 Ond\v{r}ejov, Czech
Republic}
\Email{arazimova}{sunstel.asu.cas}{cz}
\Email{kawka}{sunstel.asu.cas}{cz}
\address{$^2$Department of Physics and Space Sciences, Florida Institute of Technology, Melbourne, Florida 32901-6975, USA}
\Email{svennes}{fit}{edu} 

\markboth{SPECTROSCOPY OF HIGH PROPER MOTION DA WHITE DWARFS}{E. ARAZIMOV\'A, A. KAWKA \& S. VENNES}

\abstract{We used the revised New Luyten Two-Tenths (rNLTT) catalog to select high
proper motion white dwarf candidates. We studied the spectra of 70
hydrogen-rich (DA) white dwarfs, which were obtained at the Cerro Tololo
Inter-American Observatory (CTIO) and extracted from the Sloan Digital Sky
Survey (SDSS). We determined their effective temperature and surface gravity
by fitting their Balmer line profiles to model white dwarf spectra. Using
evolutionary mass-radius relations we determined their mass and cooling age. We also
conducted a kinematical study of the white dwarf sample and found that most
belong to the thin disk population. We have identified three magnetic white
dwarfs and estimated their surface magnetic field. Finally, we
have identified 6 white dwarfs that lie within 20 pc from the Sun.}

\section{INTRODUCTION}

White dwarf stars are the final stage of evolution for the majority of stars,
and can thus provide information about our Galaxy and its evolution. A number
of recent studies have aimed at extending our knowledge of the local population
of white dwarfs (e.g., Kawka \&~Vennes, 2006; Subasavage et al., 2007). The 
current estimate of the completeness of the local sample of white dwarfs 
($d < 20$~pc) is 80\% (Holberg et al., 2008).

Since nearby white dwarfs are likely to have a large proper motion, we have
used the revised New Luyten Two-Tenths (rNLTT) catalog (Salim \&~Gould, 2003)
to select our white dwarf candidates. We selected our candidates using the
reduced proper motion diagram ($V-J$ versus $V+5\log{\mu}$, 
Salim \&~Gould, 2002). To help distinguish between white dwarfs and cool 
subdwarfs, we used the optical-infrared colour diagram (Kawka et al., 2004).

\section{OBSERVATIONS}

\begin{figure}
\begin{center}
\includegraphics[width=6cm,height=6cm]{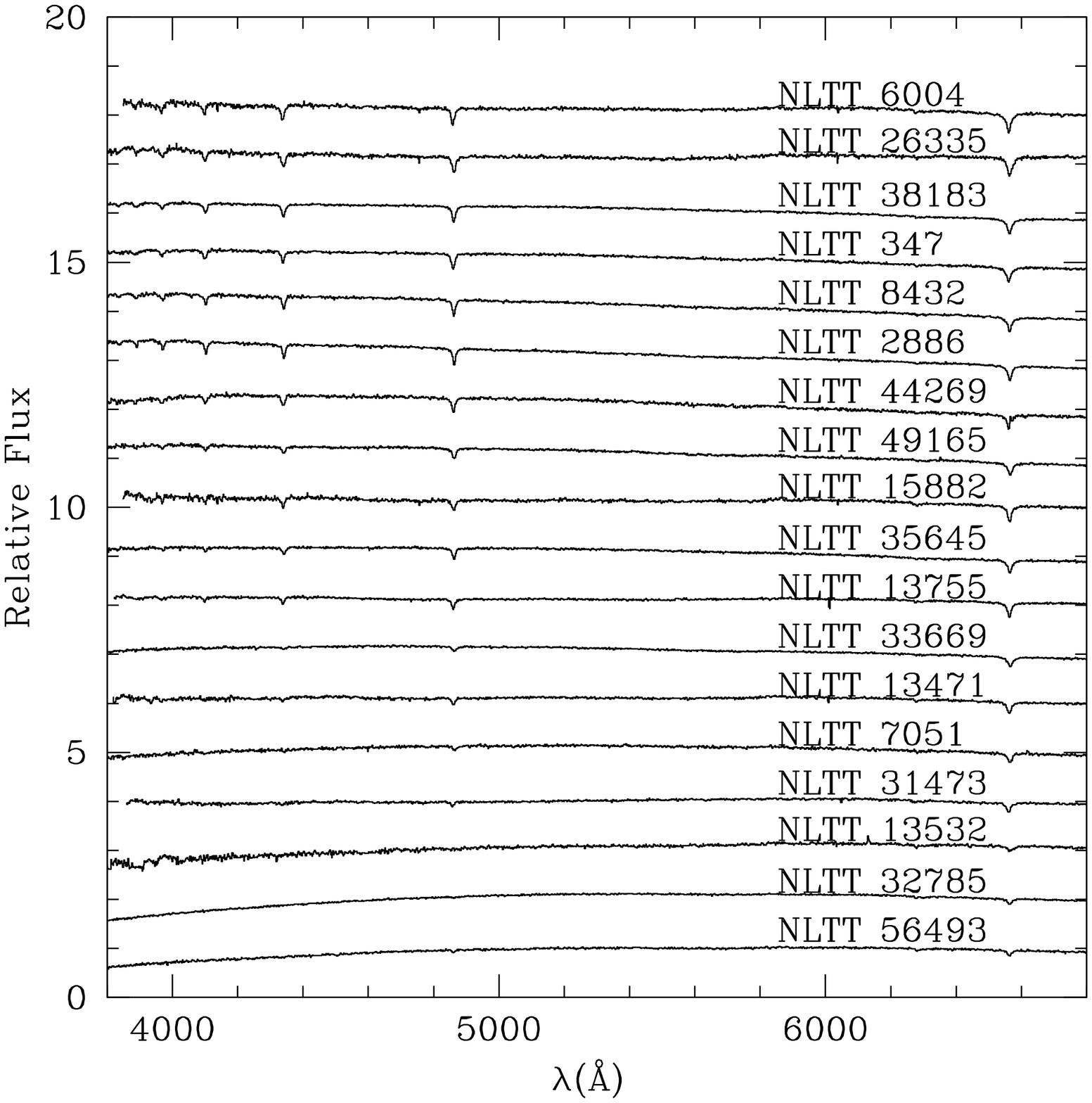}
\includegraphics[width=6cm,height=6cm]{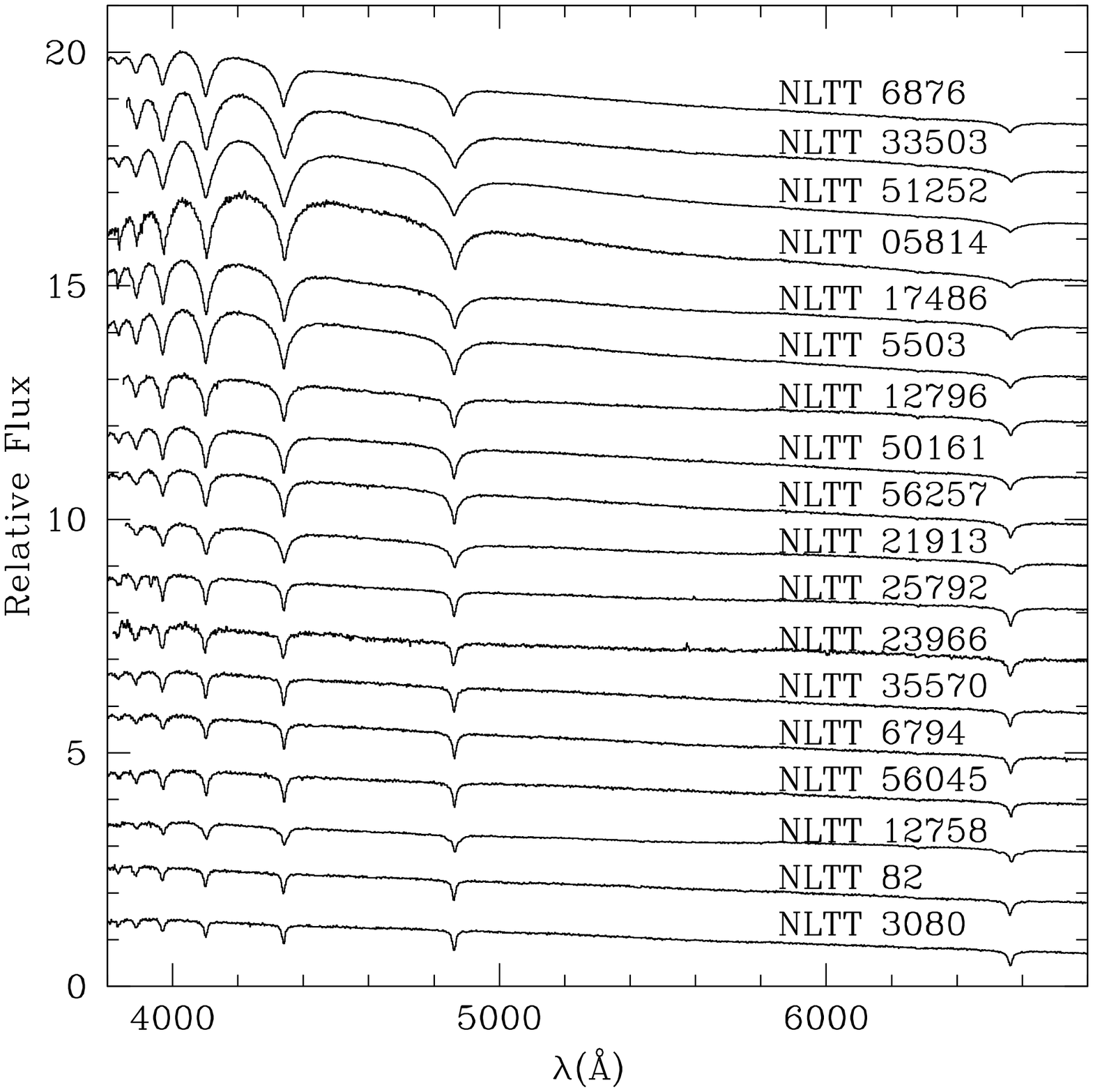}
\end{center}
\caption{Spectra of rNLTT hydrogen-rich (DA) white dwarfs obtained at CTIO. 
Effective temperature decreases from top to bottom. \label{fig_spec}}
\end{figure}

We obtained spectroscopic observations using the R-C spectrograph attached to
the 4m Blanco telescope at Cerro Tololo Inter-American Observatory (CTIO) on 
UT 2007 July 7 to 16 and 2008 February 22 to 24. We used the KPGL2 grating 
(316 lines mm$^{-1}$) with an order sorting WG360 filter. The range of the 
spectra was 3700~\AA\ to 7480~\AA\ with a dispersion of 1.99~\AA. The slitwidth 
was set to 1.5'' which resulted in a spectral resolution of $\sim$~8~\AA. 
Each night we obtained a spectrum of a flux standard, Feige~110, EG~21 or 
GD~108. Figure~\ref{fig_spec} shows the CTIO white dwarf spectra.

The current release (Data Release 6) of the Sloan Digital Sky Survey (SDSS)
spectroscopic catalog contains over 1.1 million spectra. These spectra have
a spectral coverage of 3800 \AA\ to 9200 \AA\ with a spectral resolution of
$R \sim 2000$. We cross-correlated our list of white dwarf candidates with the
SDSS and obtained 34 spectra of DA white dwarfs.

\section{ANALYSIS}

\begin{figure}
\begin{center}
\includegraphics[width=0.45\textwidth]{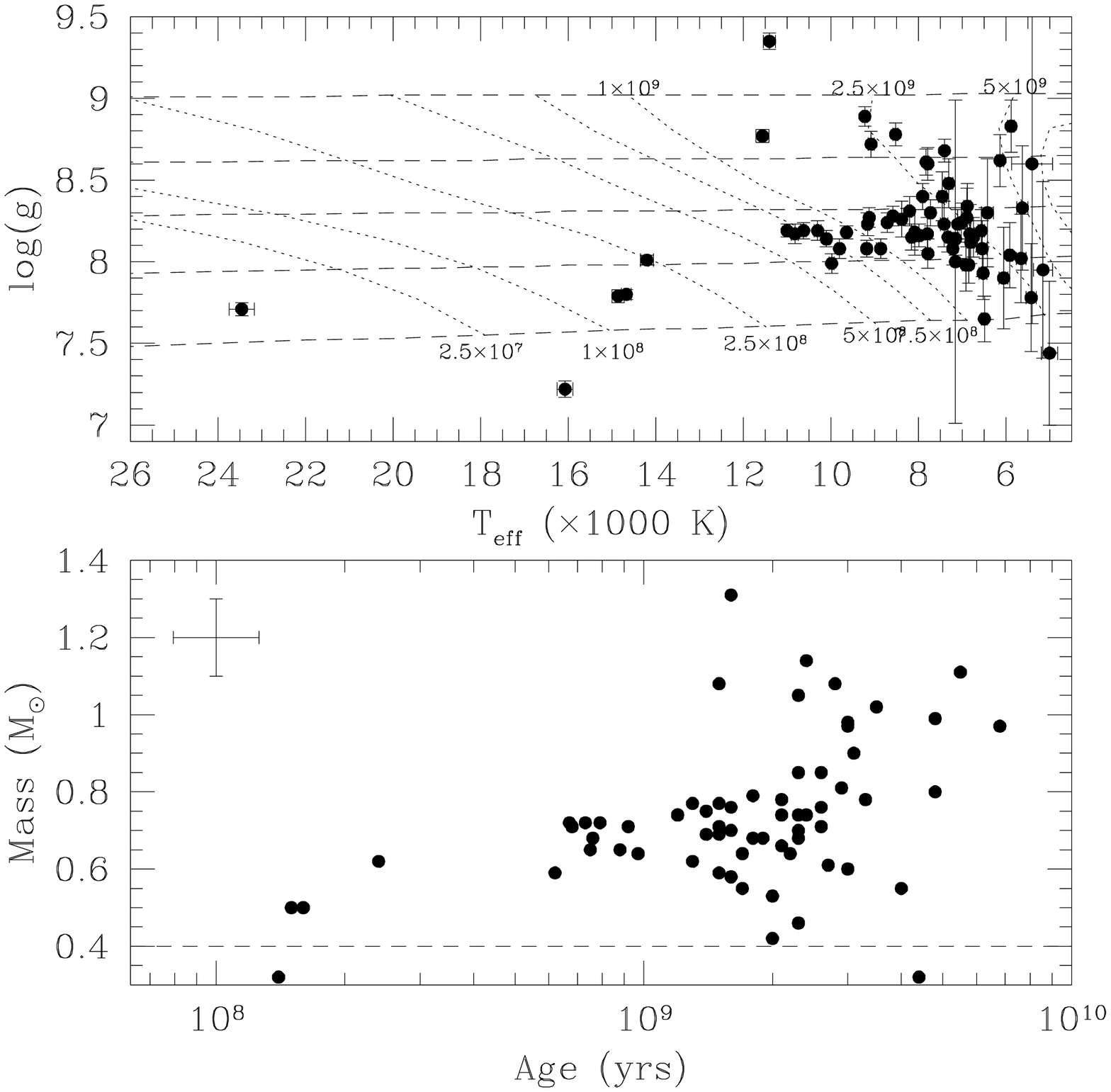}
\includegraphics[width=0.45\textwidth]{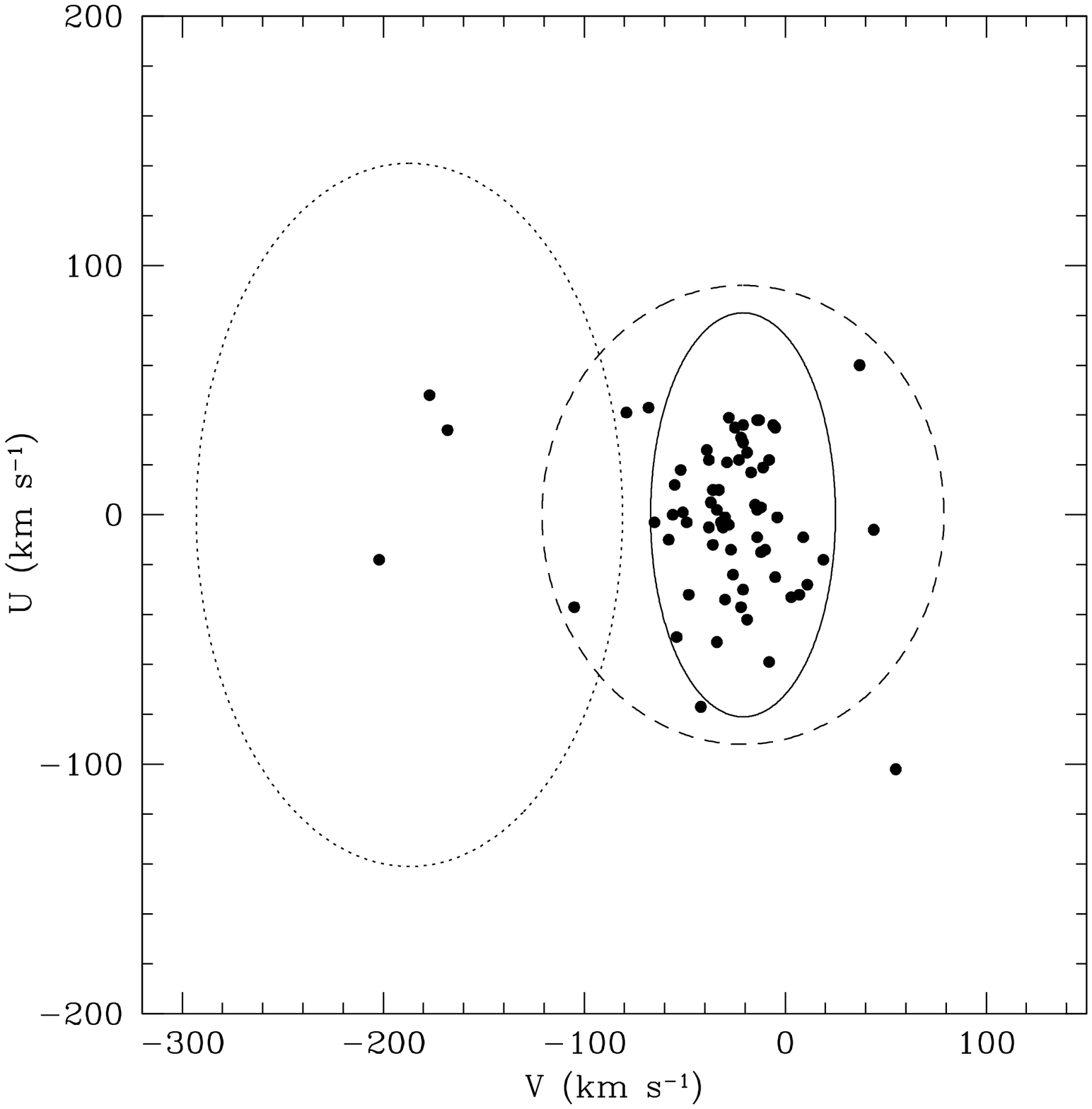}
\end{center}
\vspace{-0.5cm}
\caption{{\em Top left}: Effective temperature and surface gravity of the
rNLTT DA white dwarfs compared to the mass-radius relations of 
Benvenuto \&~Althaus (1999) and their masses and cooling ages 
({\em bottom left}). We exclude objects for which we did not determine
a surface gravity. Typical error bars for masses and ages are shown at the top 
left corner of the figure.
{\em Right:} $U$ vs. $V$ diagram showing rNLTT DA white dwarfs.
The solid and dashed lines show the $2\sigma$ velocity ellipses of the thin 
and thick disk populations, respectively. The dotted line shows the $1\sigma$ 
ellipse of the halo population (Chiba \&~Beers, 2000). \label{fig_dist}}
\end{figure}

We analyzed the white dwarf stars by fitting their spectra with a grid of
hydrogen-rich LTE models (Kawka et al., 2007 and references therein). This grid
covers effective temperatures between 4500~K and 100\, 000~K and surface
gravities from $\log{g} = 7.0$ to $9.5$. The mass and cooling age of a white 
dwarf were determined using the evolutionary mass-radius relations of Benvenuto \&~Althaus 
(1999). For masses less than 0.45~M$_{\odot}$, helium core models 
(Benvenuto \&~Althaus, 1998) were used, for masses with 
$0.45\ {\rm M}_{\odot} \le {\rm M} \le 1.2\ {\rm M}_{\odot}$ carbon/oxygen core 
models were used (Althaus \&~Benvenuto, 1997, 1998), and for masses larger than 
$1.2\ {\rm M}_{\odot}$ the mass-radius relations of Hamada \&~Salpeter (1961) 
for a carbon core were used. Figure~\ref{fig_dist} shows the effective
temperature versus the surface gravity of the white dwarfs compared to the
mass-radius relations of Benvenuto \&~Althaus (1999) and their masses and ages.

The majority of the white dwarfs in our sample are cool 
($T_{\rm eff} < 10\, 000$~K) and hence old white dwarfs, with an average 
temperature of $\sim 8300$~K. 
The average mass of our sample of white dwarfs is $M = 0.70$~M$_{\odot}$ with a
dispersion of $\sigma_M = 0.19$~M$_{\odot}$, which is slightly higher than 
the average mass of the local sample of white dwarfs 
($M = 0.665$~M$_{\odot}$ Holberg et al., 2008).
We have identified 3 new white dwarfs (NLTT 27781,
33669 and 56257) that are more massive than $1M_\odot$. All three stars are 
cooler than 10\, 000~K and the possibility that helium is contributing toward 
the broadening of the Balmer lines, and hence mimicking a massive white dwarf, 
cannot be ruled out. Parallax measurements are needed to confirm their masses. 

We have identified 3 magnetic white dwarfs in our sample, two of which are new
(NLTT~12758 and NLTT~24770). NLTT~20629 (SDSS~J085830.85+412635.1) was reported 
to be magnetic by Schmidt et al. (2003).
We have estimated
the surface magnetic field strength from the Zeeman split Balmer lines of
NLTT~12758, NLTT~20629 and NLTT~24770 to be 1.7 MG, 1.2 MG and 1.3 MG, 
respectively. All three magnetic white dwarfs are cool, 
i.e., $T_{\rm eff} < 8000$ K. 

NLTT~33108 (WD~1307+354) is a known representative of ZZ Ceti stars and it lies 
near the red edge of the ZZ Ceti instability strip. Our spectroscopically
determined parameters ($T_{\rm eff}=11\,000\pm70$~K, $\log{g} =8.19\pm0.04$)
are in agreement with previous atmospheric parameters measurements 
(e.g., $T_{\rm eff}=11\,180\pm164$~K, $\log{g}=8.15\pm0.05$, Liebert et al., 2005).

Three of our stars (NLTT 1374, 7051 and 19311) are in common proper 
motion binaries. These systems allow the gravitational redshift of the white
dwarf to be measured and hence providing another method for determining the mass.

\section{KINEMATICS}

We calculated the velocity components $U$, $V$, and $W$ using 
Johnson \&~Soderblom (1987) and assuming $v_{\rm rad} = 0$. We calculated the 
absolute magnitudes from our effective temperature and surface gravity 
measurements and then estimated the distance to the white dwarfs from the 
stars' absolute and apparent magnitudes.

Figure~\ref{fig_dist} shows the $U$ versus $V$ measurements compared to the
$2\sigma$ thin and thick disk velocity ellipses, and $1\sigma$ halo velocity
ellipse (Chiba \& Beers 2000). Figure~\ref{fig_dist} shows that most white 
dwarfs in our sample belong to the thin disk population.
Based on kinematics alone there are 4 halo candidates, but 3 of them 
(NLTT~1374, 6876 and 33503) are too hot (i.e., too young)
to belong to the old Galactic halo. Only NLTT~31473 with 
$T_{\rm eff} = 5420\pm120$~K remains a halo candidate. 

\section{SUMMARY}

We have conducted a spectroscopic study of 70 DA white dwarfs from the rNLTT
catalog, out of which half are new spectroscopically
confirmed white dwarfs. We have determined their effective
temperatures and surface gravities by fitting their Balmer line profiles to 
synthetic spectra. Using available mass-radius relations we have determined
their mass, age, distance and velocity components $U$, $V$ and $W$. We have 
discussed their membership to the different populations in the Galaxy. 
  
We have identified 6 stars that are within 20~pc of the Sun.
One of these is a known white dwarf (NLTT~19653) with a trigonometric parallax 
(Van Altena et al. 1994) that place the star at $22\pm2$~pc, which is in
reasonable agreement with our distance estimate of $18\pm2$~pc.
The remaining 5 stars (NLTT~7051, 12758, 13532, 33669 and 56257)
require trigonometric parallax measurements to confirm their distances.

\medskip
\footnotesize
E.A. is supported by grant GA \v{C}R 205/08/H005.
A.K. acknowledges support from the Centre for Theoretical Astrophysics (LC06014).

\vspace{-0.05cm}
\references
Althaus, L.G. \& Benvenuto, O.G.: 1997, \journal{ApJ}, \vol{477}, 313.

Althaus, L.G. \& Benvenuto, O.G.: 1998, \journal{MNRAS}, \vol{296}, 206.
 
Benvenuto, O.G. \& Althaus, L.G.: 1998, \journal{MNRAS}, \vol{293}, 177.

Benvenuto, O.G. \& Althaus, L.G.: 1999, \journal{MNRAS}, \vol{303}, 30.

Chiba, M. \& Beers, T.C.: 2000, \journal{AJ}, \vol{119}, 2843.

Hamada, T. \& Salpeter, E.E.: 1961, \journal{ApJ}, \vol{134}, 683.

Holberg, J.B., Sion, E.M., Oswalt, T., McCook, G.P., Foran, S. \& Subasavage, J.P.: 2008, \journal{AJ}, \vol{135}, 1225.

Johnson, D.R.H. \& Soderblom, D.R.: 1987, \journal{AJ}, \vol{93}, 864.

Kawka, A., Vennes, S. \& Thorstensen, J.R.: 2004, \journal{AJ}, \vol{127}, 1702.

Kawka, A. \& Vennes, S.: 2006, \journal{ApJ}, \vol{643}, 402.

Kawka, A., Vennes, S., Schmidt, G.D., Wickramasinghe, D.T., \& Koch, R.: 2007, \journal{ApJ}, \vol{654}, 499.

Liebert, J., Bergeron, P. \& Holberg, J.B.: 2005, \journal{ApJS}, \vol{156}, 47.

Salim, S. \& Gould, A.: 2002, \journal{ApJ}, \vol{575}, L83.

Salim, S. \& Gould, A.: 2003, \journal{ApJ}, \vol{582}, 1011.

Schmidt, G.D., Harris, H.C., Liebert, J., et al.: 2003, \journal{ApJ}, \vol{595}, 1101.

Subasavage, J.P., Henry, T.J., Bergeron, P., Dufour, P., Hambly, N.C., \& Beaulieu, T.D.: 2007, \journal{AJ}, \vol{134}, 252.

Van Altena, W.F., Lee, T.J., \& Hoffleit, E.D.: 1994, \journal{General Catalog of Trigonometric Stellar Parallaxes (4th ed.)}, New Haven, CT: Yale Univ. Obs.
\endreferences

\end{document}